# Stretchable self-tuning MRI receive coils based on liquid metal technology (LiquiTune)

**Elizaveta Motovilova[1,2,*] | Ek Tsoon Tan[2] | Victor Taracila[3] | Jana M. Vincent[3] |Thomas Grafendorfer[3]| James Shin[1] | Hollis G. Potter[2] | Fraser J. L. Robb[3] | Darryl B. Sneag[2] | Simone A. Winkler[1,*]**

[1]Department of Radiology, Weill Cornell Medicine, New York, NY, 10065, USA. [2]Department of Radiology, Hospital for Special Surgery, New York, NY, 10021, USA. [3]GE Healthcare, Ohio, USA. * Email: elm4010@med.cornell.edu; ssw4001@med.cornell.edu

**Magnetic resonance imaging systems rely on signal detection via radiofrequency coil arrays which, ideally, need to provide both bendability and form-fitting stretchability to conform to the imaging volume. However, most commercial coils are rigid and of fixed size with a substantial mean offset distance of the coil from the anatomy, which compromises the spatial resolution and diagnostic image quality as well as patient comfort. Here, we propose a soft and stretchable receive coil concept based on liquid metal and ultra-stretchable polymer that conforms closely to a desired anatomy. Moreover, its smart geometry provides a self-tuning mechanism to maintain a stable resonance frequency over a wide range of elongation levels. Theoretical analysis and numerical simulations were experimentally confirmed and demonstrated that the proposed coil withstood the unwanted frequency detuning typically observed with other stretchable coils (0.4% for the proposed coil as compared to 4% for a comparable control coil). Moreover, the signal-to-noise ratio of the proposed coil increased by up to 60% as compared to a typical, rigid, commercial coil.**

## 1 INTRODUCTION

Magnetic resonance imaging (MRI) is an indispensable technique to non-invasively depict anatomic structures and facilitate diagnosis. MRI systems routinely rely on signal detection via receive-only coil arrays, which comprise multiple surface coils arranged to cover the imaging volume[1, 2]. This multi-element configuration, compared to a single element, affords higher signal-to-noise ratio (SNR) and accommodates accelerated acquisitions [3-5]. Although, most commercial radiofrequency (RF) receive coils are rigid and inflexible, ideally a coil array needs to provide both bendability and form-fitting stretchability to accommodate various body shapes and sizes to ensure optimal SNR.

Commercial RF coils are generally built to accommodate a wide range of anatomical dimensions, which increases the mean offset distance of the coil from the anatomy and therefore reduces the available SNR. This problem becomes especially challenging when trying to use the same coils for adults as infants or small children [6]. Another challenging application is long bone imaging, as the length and circumference of the extremities vary significantly within populations [7]. Some commercial arrays provide limited mechanical flexibility, with portions that can be partially folded around the area of interest; this improves coupling between the imaged volume and the coil and affords a slightly higher filling factor, thus improving RF receive efficiency. However, these designs are bulky and limited in their flexibility to a single direction. The only known commercial coil that has a considerably high degree of flexibility, yet is not stretchable, uses AIR$^{TM}$ Technology (GE Healthcare, Inc.) [8]. AIR$^{TM}$ Technology enables creation of blanket-like RF coil arrays that have improved flexibility and conformance to different anatomies. Such flexibility and form-fitting adaptability is achieved by means of a proprietary process that yields low reactance and low-loss conductors while being lightweight, flexible, and durable [9-12]. However, due to the proprietary nature of AIR$^{TM}$ Coils the stretchability of technology is not known.

Highly flexible receive RF coils have been the focus of research for many years offering rigid-bendable [13-21] and geometrically adjustable [22-25] solutions. Recent developments in high impedance coaxial coils have demonstrated high flexibility and form-fitting adaptability while also providing good element isolation [19]. However, the individual coil diameter cannot be chosen freely as it is dictated by the desired resonance frequency and properties of the coaxial cable. Multi-turn, multi-gap cable coils provide greater degrees of freedom in terms of coil size [20, 26]. However, commercially available coaxial cables have a limited and discrete set of impedances leading to discrete values of achievable coil diameters. Although the goal of high flexibility in RF coils has been partially solved by the aforementioned designs, full adaptability arguably requires coils to be stretchable as well -- a concern, which to date has not been fully addressed.

To achieve stretchability and flexibility, new materials and concepts are required. The use of liquid metal (mercury) as the coil conductor [27] was suggested as early as 1986. Mercury was contained in a flexible plastic tubing to achieve more flexibility, and to accommodate coil positioning closer to the imaging region [27]. However, mercury's toxicity, and its tendency to assume a spherical shape due to surface tension, limited its use.

Gallium-based liquid metals with low vapor pressure render a safer substitute. Gallium and its alloys have already found many applications in biomedical fields – for example, as restorative materials in dentistry, as tumor imaging and tumor growth suppression agents, and for the treatment of certain cancers [28]. The fluidity, low viscosity, and low melting point of gallium-based alloys make them easy to handle with a syringe at room temperature. When in contact with air, an oxide layer forms on the surface, which prevents the inner metal from further oxidization, while also allowing liquid metals to adhere to surfaces and adopt useful shapes. The electrical resistivity of gallium is 13.6·10⁻⁸ Ω·m, which is higher than copper (1.68·10⁻⁸ Ω·m) but lower than liquid metal mercury (9.8·10⁻⁷ Ω·m)[29]. However, as was demonstrated recently in [30], MRI coils made with alternate conductive materials (such as aluminum, liquid metal, conductive polymer, and braided conductors) can achieve SNR levels that may exceed intuitive expectations [31] despite their higher resistance compared to commonly used copper-based coils. As long as the thickness of the conductor is greater than several skin depths, even a significant change in coil resistance will only marginally affect the SNR.

Recent materials science developments allow for the creation of extremely flexible polymers that can undergo significant degrees of stretching, making these polymers feasible as substrates to realize ultra-stretchable conductors with liquid metal encapsulated inside [32].

To date, only a few studies have described applying liquid metal technology to stretchable MRI RF coil design [33-37]. In [34], the authors directly deposited liquid metal on a stretchable, neoprene fabric to create a flexible knee coil. In [37], a stretchable silicone tube with encapsulated liquid metal was used to build a four-element array, also for the knee. These stretchable coils, however, suffer from resonance detuning. When a coil is stretched, its length and thus its inductance is increased, which leads to a resonance frequency shift. Proposals to mitigate this effect include wide-band matching [36] and automatic tune-match circuitry[35, 38]. However, such designs require additional fixed and inelastic electrical components and circuitry that can dramatically increase complexity and thus make reliability and implementation more complex.

In this study, we describe a method of fabricating soft and stretchable RF receive coils based on liquid metal and ultra-stretchable polymer with a smart geometry design to provide autotuning capability and mitigate the resonance frequency shift, without using additional tuning circuitry. This proof-of-concept work includes theoretical analysis and numerical simulations, which are subsequently confirmed by bench experiments and MRI tests of a single element RF receive coil at 3T in vitro and in vivo.

## 2    RESULTS

**Theoretical analysis.** The proposed coil geometry comprises a single interdigital capacitor and a rectangular loop (Fig.1a). The liquid metal conducting traces are embedded in a stretchable polymer matrix. To understand the resonance behavior of such a coil under stretch and theoretically describe it in a first order approximation, we developed an analytical circuit model where coil parameters such as trace width, digit length, width and spacing are functions of the degree of stretch. Approximate analytical formulas for coil inductance and capacitance are then used to calculate how these values change with parameter variation (equations (1)-(4)). Assuming linear stretching along the x-axis, the inductance increases approximately linearly with stretching, while the capacitance of the interdigital capacitor decreases (Fig. 1(b)). As the resonance frequency depends on the product of inductance and capacitance, $f_0 = 1/(2\pi\sqrt{L \cdot C})$, this analysis indicates that it is possible to compensate for the frequency shift due to the inductance changes with an optimized capacitor design. Fig. 1(c) compares the changes in resonance frequency of the control stretchable coil (blue color) with a fixed value capacitor and that of the proposed coil (red color) with a single stretchable interdigital capacitor. The smart geometry of the proposed coil helps to reduce the resonance frequency shift. However, the non-linear change in capacitance cannot fully compensate the linear change of inductance. That is why we still observed frequency shift with stretching. However, the frequency shift of the proposed coil is less significant (33% less) compared to the control coil that has a fixed value capacitor. All coil and interdigital capacitor parameters can be dynamically changed in the analytical circuit model, and the optimal range of parameter values can be identified that suit the desired coil requirements. Although these parameter values provide a reasonable starting point, the accuracy of the analytical model is limited to the first order approximation of the nearest-neighbor interactions, and thus further rigorous numerical simulations were performed as shown in the next section.

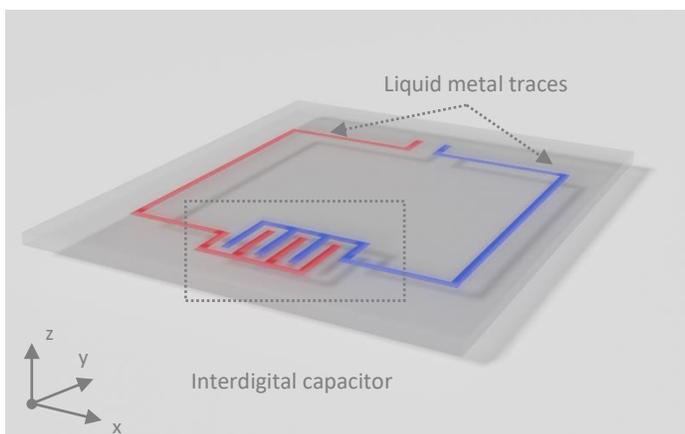

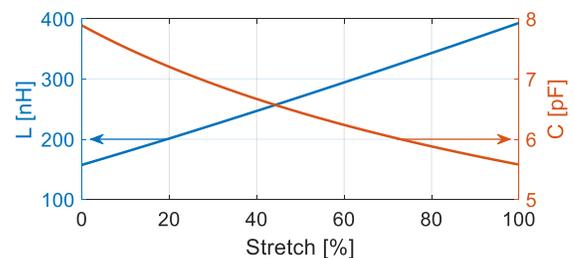

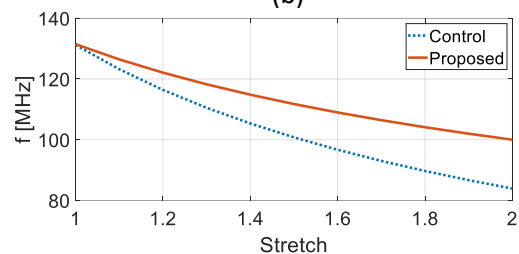

*Fig.1: (a) Schematic of the proposed RF coil element. (b) Inductance and capacitance changes with stretch. (c) Resonance frequency changes with stretch of the proposed (red) and control (blue) coils indicating frequency compensation ability of the stretchable interdigital capacitor.*

**Numerical simulations**

A 3D-model of the proposed coil with an interdigital capacitor was developed (Fig.2(a)) and full wave electromagnetic simulations were performed. The parameters of the interdigital capacitor were optimized such that the resonance frequency of the coil has minimal shift with stretching from 0% to 30%. For comparison purposes, a control coil (Fig.2(b)) of the same dimensions with a fixed value capacitor instead was modelled as well to serve as a reference coil similar to those utilized in previously-proposed liquid metal coil designs [34, 39].

Input impedance $S_{11}$ changes with respect to coil stretching from 0% to 30% were simulated for the control and proposed coils, and the results are summarized in Fig.3(c). Initially (at 0% stretch), both coils were tuned to 128MHz (the operation frequency of a 3T clinical scanner) and matched to $50\Omega$; no further tuning/matching was thereafter performed. With stretching applied, the input impedance $S_{11}$ of the control coil (blue curves) predictably linearly shifts towards lower frequencies. Alternatively, the input impedance of the proposed coil (red curves) fluctuates near the initial resonance frequency by first shifting to higher frequencies (for up to 15% stretch) and then returns to lower frequencies (for ≥20%). As expected, the 3D numerical model yields improved accuracy and illustrates the non-linear behavior of the coil model as compared to the first order approximation in the theoretical model. This is because the 3D numerical model takes into account 1) higher-order interactions (between non-near neighboring digits of the capacitor), 2) material properties and losses, and 3) coil loading, to realistically represent the actual setup.

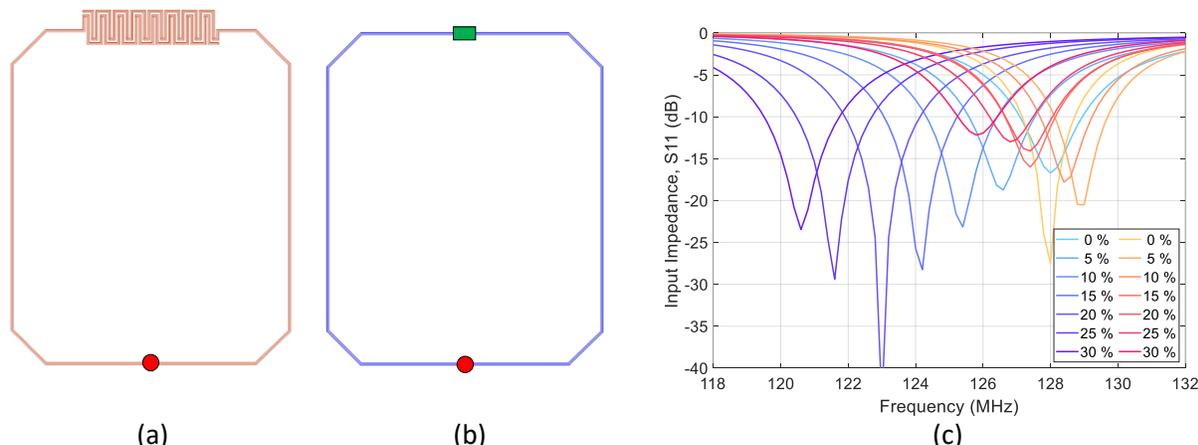

(a)  (b)  (c)

*Fig.2. Simulation models of the (a) proposed coil with stretchable interdigital capacitor and (b) control coil with fixed value capacitor (green rectangle). (c) Simulated input impedance ($S_{11}$) change with stretch for the proposed (red) and control (blue) coils showing frequency stability of the proposed coil.*

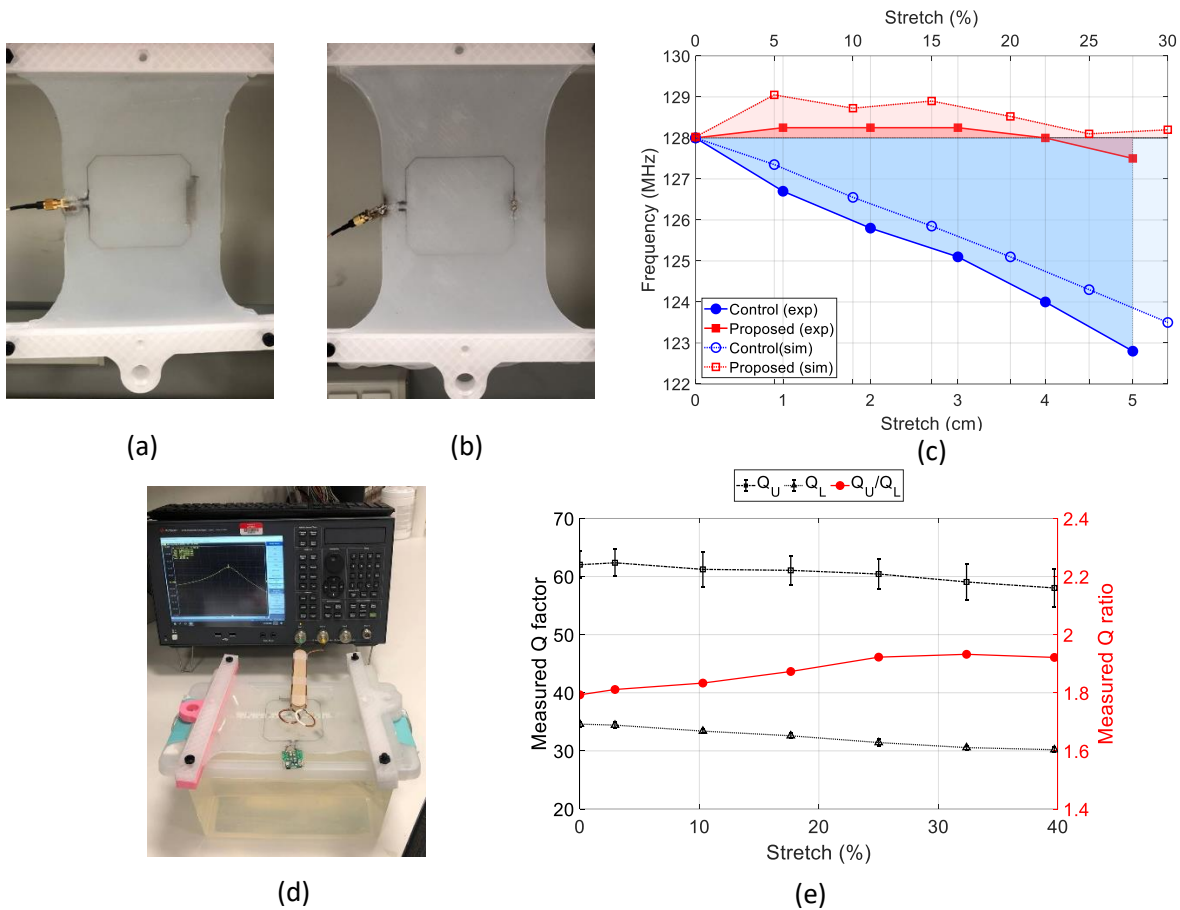

Fig.3: Fabricated prototypes of the (a) proposed and (b) control coils attached to a stretch testing rig. (c) Simulated and measured frequency shift with strech for the proposed (red) and control (blue) coils, indicating improved resonance frequency stability with stretching for the proposed coil. (d) Proposed coil prototype placed on a cylindrical phantom with a tape measure for stretching measurement. (e) Measured loaded $Q_L$ and unloaded $Q_U$ quality factors (black) of the proposed coil and their ratio $Q_U/Q_L$ (red), indicating that the coil losses are dominated by the losses in the phantom.

Next, the proposed coil was modelled with a homogeneous rectangular phantom that represents average tissue properties. Sensitivity ($B_1^-$ field normalized to 1W input power) profiles at the central axial cross-section were simulated at several stretching levels as illustrated in Fig.4(a)-(e). This cross-section depicts how coil sensitivity is relatively unaffected by stretching. Although sensitivity at the center decreases slightly, the coil is able to cover a larger area with stretching while maintaining an excellent performance (7.5±0.6µT at the surface of the phantom).

**Bench measurements**

Fig.3 shows a portion of the bench measurement setup, where two coil prototypes, (a) proposed and (b) control, are connected to a 3D-printed unidirectional stretch testing rig. The coils are connected to a vector network analyzer through an L-shaped tune/match network and the input impedance $S_{11}$ changes with stretching are reordered. To more clearly portray the coil performance, the resonance frequency is plotted against the degree of stretch, as shown in Fig.3(c), where the curves indicate the actual frequency change of the control (blue) and proposed (red) coils, and the shaded areas help to visualize the total frequency shift. Fig.3(c) shows both simulated (line with empty markers) and measured (lines with solid markers) changes in resonance frequency with stretching, illustrating that the measured data agree well with the simulated results and demonstrating an improved frequency stability of the proposed coil. In particular, at the measured maximum stretch of 5cm (corresponding to ~27% stretching), the frequency shift for the proposed coil is only 0.5MHz (0.4%) as compared to >5MHz (>4%) for the control coil, which indicates a 10-fold reduction in frequency shift. Fig.3(d) shows the proposed coil placed on a rectangular phantom ($\varepsilon_r = 78, \sigma = 0.8$ S/m) with a tape measure to track dimension changes with stretching. The quality factor of the proposed coil was measured using a pickup probe when the coil was loaded with this phantom $Q_L$ and without the phantom $Q_L$. The quality ratio $Q_{ratio} = Q_U / Q_L$ is around two, which means that the majority of coil losses are coming from the phantom and not from the coil itself. Although, the $Q_{ratio}$ is not very high as compared to traditional copper-base RF coil (~10) due to the higher resistivity of liquid metal, it is comparable to the values reported in the literature for similar stretchable coils [34, 37, 39, 40].

**In-vitro imaging**

Fig.4 (f)-(j) show the SNR maps obtained with the proposed coil when the element is positioned on a homogeneous rectangular phantom. In this first prototype, the Ecoflex® material used for the stretchable polymer matrix and the liquid metal appear bright on the image; thus, the two points of hyperintensity correspond to the edges of the liquid metal coil, and it is convenient to track the extension of the coil. The coil was gradually stretched from 0% (Fig.4 (f)) to 50% (Fig.4(j)). The supplementary video V2 demonstrates the measured SNR as the coil stretches. The coil element was optimized for stretching levels up to 30% (Fig.4(h)-(i)) and demonstrates maintained SNR when stretched within the limits ($SNR = 516 \pm 46$, at the surface of the phantom, which translates to SNR variations of less than 9%). Interestingly, it continues to provide useful SNR even beyond its designed limits Fig.4(j), at levels comparable to a rigid, commercial coil. As the SNR of surface coils is a function of coil dimensions [41], it is expected to decrease for larger coil areas. This means that, all other parameter being equal, the SNR of a stretchable coil will decrease with the degree of stretch. Therefore, to fairly compare the coil performance at various stretching levels, its SNR maps were normalized ($SNR^*$) to the corresponding coil size at each stretch level as follows: $SNR^* = SNR_i/A_i$, where $SNR_i$ is the measured SNR map and $A_i$ is the corresponding coil area at a particular stretch level $i$. These normalized maps (Fig.4(k)-(o)) clearly illustrate stable $SNR^*$ performance over the designed stretching levels (up to 30%). When the normalized $SNR^*$ is measured through the center of the coil into the phantom (Fig.4(p)), the measured values are $SNR^* = 561 \pm 12$, which indicates a variation of only 2%.

**In-vivo imaging**

Fig.5 shows sagittal fast spin echo images of a healthy knee acquired with the proposed single element coil when the coil element is (a) unstretched and (b) 15% stretched. The images

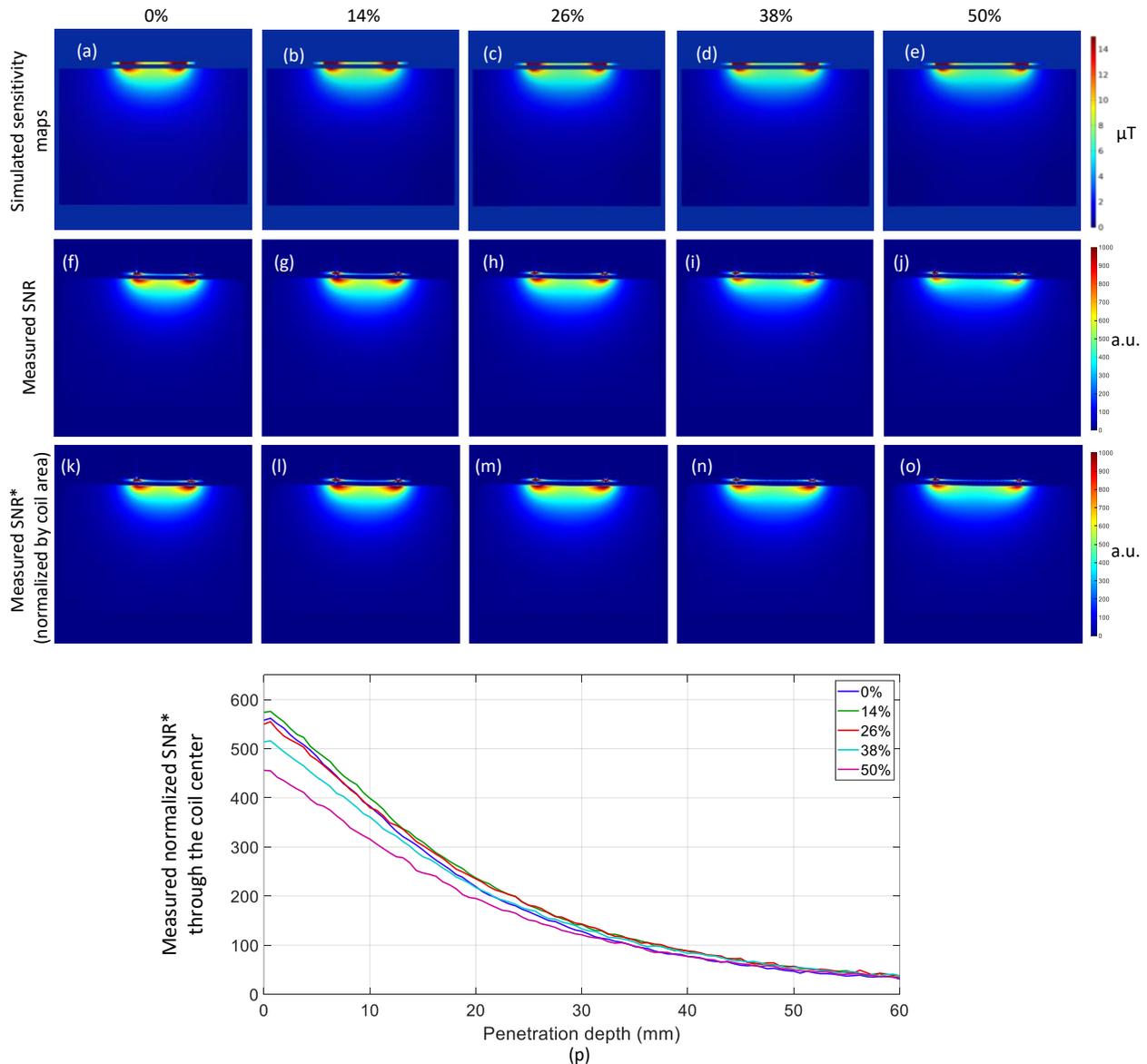

*Fig.4. Phantom/in vitro stretching tests with the proposed coil. (a)-(e) Simulated sensitivity (B1- field) profiles (normalized to 1W input power) and (f)-(j) measured SNR maps at the central axial slice with the numbers on top indicating the percentage of coil stretch. (k)-(o) Measured SNR maps normalized to the coil area demonstrating relative SNR stability with stretching. (p) Normalized SNR\* measured through the center of the coil into the phantom.*

are compared to their counterpart acquired using (c) a dedicated 8-channel knee coil. Only a portion of the knee is visible on the first two images (a)-(b) as they were acquired with a single channel surface coil, while the whole knee can be seen in (c) as it was acquired using a volume multi-channel coil. An SNR comparison using the same region of interest (ROI) highlighted as the dashed yellow rectangle in each image highlights the advantage of the proposed coil. The SNR values measured were

(a) 282, (b) 288, and (c) 179, which corresponds to a 60% SNR increase of the proposed coil compared to the dedicated knee coil array. Fig.5(d)-(f) show the corresponding SNR maps of the same saggital slices demonstrating the imporved SNR values of the proposed coil compared to the dedicated knee coil. This significant SNR improvement in the single-element case alone is achieved due to the conformal design of the stretchable coil. We expect to even further improve SNR with an

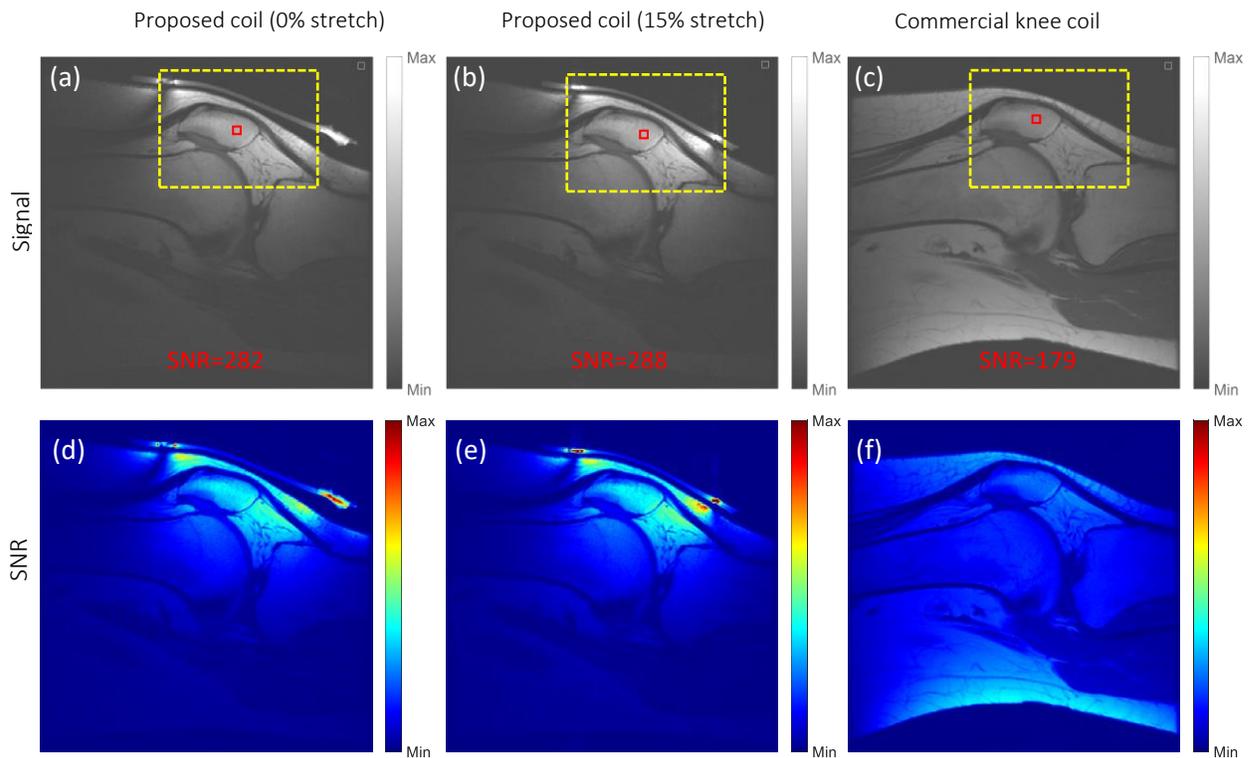

Proposed coil (0% stretch)    Proposed coil (15% stretch)    Commercial knee coil

*Fig.5.(a)-(c) In vivo knee imaging using different coils showing improved SNR of the proposed coil when comparing the same ROI (yellow box). Numbers in red indicate SNR values in the selected ROI (red square). (d)-(f) The corresponding SNR maps demonstrating improved SNR*

optimized dedicated multi-channel array. As a first proof of concept, this image demonstrates feasibility of the proposed concept -- in particular, the form-fitting conformability of the coil as well as its stretchability and frequency stability, while providing increased SNR as compared to commercial state-of-the-art RF coils.

## 3    DISCUSSION

In this work, we present a novel stretchable MRI receive coil design with self-tuning capability, applied to a 3T MRI system to demonstrate feasibility. Liquid metal embedded in a soft and highly elastic polymer matrix allows one to freely position and stretch the coil to conform to various anatomies. In addition, the smart geometry of the stretchable interdigital capacitor maintains frequency stability within 0.5% over a large range (up to 30%) of stretching conditions. We observe a dramatic SNR improvement (31%), even in this limited single-channel proof-of-concept prototype, as compared to the existing rigid commercial array.

The coil, it can be applied to a plethora of complex anatomies, such as the breast/chest wall, perineum/groin, neck/shoulder area and digits for improved SNR and sensitivity. The frequency stability under various coil stretches also allows for dynamic and kinematic imaging of various body/joint positions, such as the knee or wrist, at a wide range of flexion angles, which is not possible with current rigid and semi-flexible commercial coils.

This initial prototype is designed to provide frequency stability under a unidirectional stretching. A two-dimensional prototype is a straightforward extension of the proposed concept by addition of capacitors in the second dimension and is the subject of current and future work.

In this work, the first proof-of-concept, self-tuning stretchable RF coil is demonstrated in a single channel design. Future work will tackle a multi-channel array design, where decoupling strategies, such as geometric decoupling, will be implemented.

In this purely technical work, we compared the proposed coil to a state-of-the-art commercial knee coil. To the attentive reader, a comparison to other flexible concepts might be of value. We would like to point out that the main aspect of our proposed work is its inherent stretchability as opposed to the flexibility/bendability, which is almost a side benefit of the stretchability. While GE AIR$^{TM}$ coils, screen-printed designs, and cable coils provide flexibility as their shape can be changed, the total coverage area remains the same – there is hence no inherent stretchability for these designs. In particular, the area of the proposed design changes from 49cm$^2$ to 61cm$^2$ when stretching from 0% to 50%, while also maintaining stable SNR. A comparison to the more established, flexible-only designs, will be the goal of a more clinical investigation once the LiquiTune technology is built in the form of an array. This will include literature-based comparisons to flexible coil designs, such as GE AIR$^{TM}$ technology [8], cable coils [21], screen-printed arrays [6, 16]and other liquid metal-based designs [37, 39].

Minor artifacts are observed in the anterior/posterior direction and are attributed to the polymer and/or liquid metal composition, which will be optimized in a future prototype. Moreover, manganese (Mn) or gadolinium (Gd) components can be added to the Ecoflex material used for the stretchable polymer matrix to increase relaxivity and hence reduce signal in MR images.

It was shown in [42] that although addition of silver microparticles reduces the conductivity of gallium, gold microparticles can slightly increase the conductivity of liquid metal. Optimizing the liquid metal alloy composition in this manner may improve the conductivity of liquid metal which reduces the coil losses, improves the Q factor, and effectively increases the SNR of the RF coil further.

Gallium alloys can remain stable within the polymer matrix; however, there is a small chance of liquid metal spilling if there is a tear in the polymer. As an alternative, other flexible conductive materials can be used, such as a conductive elastomer with silver nanoparticles used in [39]. Although the conductivity of such materials is inherently lower compared to metal, coils based on such materials can provide comparative Q-factor and SNR values [39]. Future work will discuss appropriate sealing methods to further minimize risk of damage or spillage, as well as to facilitate practical requirements in the clinic, such as sanitation.

Although the permittivity of Ecoflex material is relatively low ($\varepsilon_r = 2.8$), it can be increased by mixing the uncured polymer with a high permittivity material such as barium titanate, BaTiO$_3$(BTO) nanoparticles [43]. This will affect the value of the interdigital capacitor and provide another degree for design optimization.

## 4  CONCLUSIONS

In this work, we propose a novel concept of stretchable and self-tunable RF coil for MR imaging based on liquid metal integrated within a soft polymer matrix. We demonstrate resonance frequency stability within 0.4%, SNR fluctuations within less than 9% *in vitro*, and more than 60% increased SNR *in vivo* compared to state-of-the-art commercial coils.

The proposed self-tuning stretchable coil design provides a pathway towards highly flexible and conformal MR coils that provide increased SNR, facilitate dynamic/kinematic imaging, and offer greater patient comfort. This is in line with the general trend in modern technology where comfort-centered, lightweight and wearable devices are at the forefront of novel developments. MRI technology is bound to follow suit with the help of safe liquid metals and ultra-stretchable polymers.

## 5 METHODS

**Theoretical analysis.** The approximate resonance behavior of the proposed RF coil element can be theoretically analyzed using a simplified circuit model. When the dimensions of a coil loop element change, its resonance frequency changes accordingly. The frequency shift can be understood through the resonance equation,

$$f_0 = 1/(2\pi\sqrt{L \cdot C}), \qquad (1)$$

where $f_0$ is the resonance frequency, and $L$ and $C$ are the total inductance and capacitance of the coil, respectively. The coil inductance is proportional to the coil conductor length, which in turn indicates that if a coil is stretched, the coil inductance is increased and the resonance frequency is decreased. The total self-inductance of a loop coil under a stretch $\alpha$ changes approximately as follows

$$L = \alpha L_0 + \mu_0\mu_r(3D/4)\alpha\ln(\alpha), \qquad (2)$$

with the inductance $L_0$ in the original unstretched state as given by [44]

$$L_0 = \mu_0\mu_r(D/2)\{\ln(8D/d) - 2\}, \qquad (3)$$

where $\mu_0 = 4\pi10^{-7}$H/m is the vacuum permeability, $\mu_r$ is the material permeability, $D$ is the loop diameter and $d$ is the wire diameter. To mitigate the frequency shift, we use an interdigital capacitor that decreases its capacitance under stretch. It was shown in [45] that when a stretch $\alpha$ is applied perpendicular to the fingers (digits) of the interdigital capacitor, the total capacitance is changing inversely proportionally to the square root of the stretch $C = \dfrac{C_0}{\sqrt{\alpha}}$, where $C_0$ is the initial capacitance of the interdigital structure. The total capacitance of an interdigital capacitor can be approximated as follows,

$$C_0 = (N-3)C_I/2 + 2\frac{C_I C_E}{(C_I + C_E)}, \qquad (4)$$

with $C_I = \varepsilon_0\varepsilon_r p\dfrac{K(k_{I\infty})}{K(k'_{I\infty})}$, $C_E = \varepsilon_0\varepsilon_r p\dfrac{K(k_{E\infty})}{K(k'_{E\infty})}$, where $N$ is the number of digits, $\varepsilon_0$ is the vacuum permittivity, $\varepsilon_r$ is the host material permittivity, $p$ is the digit length, $K(k)$ is the complete elliptical integral of the first kind with modulus $k$, $k' = \sqrt{1 - k^2}$ is the complementary modulus, and $k_{I\infty} = sin(\pi\eta/2)$, $k_{E\infty} = \dfrac{2\sqrt{\eta}}{1+\eta}$, $\eta = \dfrac{w}{w+g}$ [46]. Thus, both capacitance and inductance can be expressed through the coil parameters $D, d, w, g, p$, and the resonance frequency of the coil can be calculated with respect to these parameters and various stretching levels, $\alpha$. Although the complete coil behavior under stretch is more complicated and requires rigorous numerical modeling, this simplified theoretical analysis demonstrates in first order approximation how the integration of a stretchable interdigital capacitor into an RF loop coil can help to reduce coil detuning under stretch.

**Numerical simulations.** Once approximate parameter values are obtained from the theoretical analysis, full-wave numerical simulations are required to perform exact parametric optimization of all dimensions. To this goal, a single-element coil, as shown in Fig.2, was modeled using COMSOL Multiphysics [47] and loaded with a homogeneous ($\varepsilon_r = 78$, $\sigma = 0.46$ S/m) cylindrical ($R_{ph} = 5$cm, $L_{ph} = 11$cm) phantom (not shown in the figure for simplicity). The coil element comprises a single interdigital capacitor and a rectangular $6cm \times 7cm$ loop with its terminals connected to a lumped port (indicated as red dot). Liquid metal conductors are encapsulated in a polymer matrix ($\varepsilon_r = 2.8$, $\tan(\delta) = 0.02$) and modeled as perfect electric conductors. Exact parameters of the interdigital capacitor are obtained via a rigorous parameter optimization to maintain a relative frequency stability for stretching levels from 0% to 30% yielding the following values: number of digits $N = 8$, digit length $b = 7mm$, inter-digit spacing $g = 0.5mm$ and conductor width $w = 0.5mm$. The minimum conductor width was chosen based on the manufacturing limitations (specification of the 3D printer). The minimum conductor dimension is much larger than the skin depth of the liquid metal (23.9µm). The frequency stability of the coil was evaluated for different degrees of unidirectional stretch along the *x*-axis, and the capacitor parameters are optimized to minimize the frequency shift. For comparison purposes, a second, geometrically identical, prototype of the coil was modeled where a fixed value capacitor was used instead of the flexible interdigital one. Both coils were tuned to 128MHz (the operating frequency of a 3T scanner) and matched to 50 Ohm at their unstretched state. No more retuning/rematching were done after this for all consecutive stretching cases. For the $B_1$ field sensitivity maps, a homogeneous ($\varepsilon_r = 78$, $\sigma = 0.46$ S/m) rectangular ($33cm \times 22cm \times 16cm$) phantom was used to clearly illustrate sensitivity changes when the coil undergoes a unidirectional stretch. The coil was fed using a uniform port excitation with an input power of 1W. The field maps were acquired at the central axial slice going through the middle of the coil. The sensitivity at the surface of the coil was measured at the coil center for each stretch position.

**Coil fabrication.** The RF coil was constructed by filling microfluidic channels made of stretchable insulating silicone Ecoflex™[48] with liquid metal (Gallium–Indium eutectic, EGaIn [49]). The microfluidic channels of square cross-section of width $w = 0.5mm$ (corresponding to an approximate wire gauge of AWG23) were made by bonding two silicone layers of 1.5mm thickness each and total size of $10cm \times 11cm$, where the top layer contains microfluidic channel features and the bottom layer is featureless. Detailed step-by-step instructions with illustrations of the coil manufacturing process are shown in the Supplementary Materials (Figure S1 and the corresponding text). The plastic molds that contain microfluidic channel features of the desired coil geometry and the sealing featureless top layer are fabricated using a high-resolution 3D printer (Prusa i3 MK3S, using a 25µm nozzle and a 50µm layer height) with polylactic acid (PLA) material. Liquid metal is injected into the embedded microchannels using a needle and syringe to form conducting traces. The Supplementary video V1 demonstrates the stretchability of the fabricated coil.

Copper wires are inserted at the terminals of the microchannels and connected to a feed-board. For the stretching experiments, the coil was tuned and matched with an L-shape tuning/matching circuit consisting of a shunt (10pF) and a series (6.8pF) capacitor. For the imaging experiments, the coil was connected to a custom-made integrated feed board of dimensions 22 × 22 × 10 mm that contains the matching and decoupling circuit, preamplifier, and a balun, as shown in circuit schematic in Figure S3. For comparison purposes, a second, geometrically identical, prototype of the coil was fabricated, where a fixed value capacitor was used instead of an interdigital capacitor. This variant is called "control coil" throughout the paper and was used to quantify self-tuning performance.

**Bench tests.** The two coil prototypes (self-tuning and control) were tested on the bench to measure their $S$-parameter changes under stretch using a vector network analyzer (Keysight E5071C). For this purpose, a simple stretching rig was built to facilitate controlled unidirectional coil elongation. The coils were linearly stretched with a step size of 1cm and the $S_{11}$ values were recorded. The experiment was repeated three times and the mean value was reported. The quality factor $Q$ of the proposed coil was measured using a pick-up coil as the ratio of the center frequency $\omega_0$ to the 3dB bandwidth $\Delta\omega$ of the insertion loss ($S_{21}$) with $Q_L$ and without $Q_U$ a phantom. The phantom ($23cm \times 14cm \times 13cm$) was made in house using the recipe from [50, 51] to represent dielectric properties of an average human body tissue ($\varepsilon_r = 78, \sigma = 0.8$ S/m).

**Imaging.** All experiments were performed on a GE Discovery MR750 3T system. For the in-vitro experiments, the coil was loaded with a standard homogeneous rectangular ($33cm \times 22cm \times 16cm$) phantom. Phantom images were acquired using a fast spin echo sequence with the following parameters: $TR = 315$ms, $TE = 68.7$ms, $1mm \times 1mm \times 3mm$, $NEX = 1$. For in-vivo experiments, a healthy volunteer (F, Age=29) was used. In-vivo images were acquired using the same fast spin echo sequence. The SNR values were calculated according to Method 4 described in the NEMA Standards Publication MS 1-2008 (R2014, R2020), as the ratio of the mean pixel value of signal within the specified ROI divided by the standard deviation of the noise calculated in the background region of the image, well removed from the phantom and any visible artifacts. The SNR maps were calculated by dividing the entire image by the standard deviation of the noise calculated as described above. The study was performed with ethical approval from the Weill Cornell Medicine Institutional Review Board and in accordance with all applicable regulations. Informed consent was obtained from the volunteer.

## REFERENCES

[1] Roemer PB, Edelstein WA, Hayes CE, Souza SP, Mueller OM. The NMR phased array. Magnetic resonance in medicine 1990;16(2):192-225.

[2] Vaughan JT, Griffiths JR. RF coils for MRI. John Wiley & Sons; 2012.

[3] Sodickson DK, Manning WJ. Simultaneous acquisition of spatial harmonics (SMASH): fast imaging with radiofrequency coil arrays. Magnetic resonance in medicine 1997;38(4):591-603.

[4] Pruessmann KP, Weiger M, Scheidegger MB, Boesiger P. SENSE: sensitivity encoding for fast MRI. Magnetic Resonance in Medicine: An Official Journal of the International Society for Magnetic Resonance in Medicine 1999;42(5):952-62.

[5] Griswold MA, Jakob PM, Heidemann RM, Nittka M, Jellus V, Wang J, et al. Generalized autocalibrating partially parallel acquisitions (GRAPPA). Magnetic Resonance in Medicine: An Official Journal of the International Society for Magnetic Resonance in Medicine 2002;47(6):1202-10.

[6] Winkler SA, Corea J, Lechêne B, O'Brien K, Bonanni JR, Chaudhari A, et al. Evaluation of a flexible 12-channel screen-printed pediatric MRI coil. Radiology 2019;291(1):180-5.

[7] Fryar C, Gu Q, Ogden C, Flegal K. Anthropometric reference data for children and adults: UnitedStates, 2011–2014. In: Statistics NCfH, ed. 3 (39). 2016.

[8] AIR Technology; Available from: https://www.gehealthcare.com/products/magnetic-resonance-imaging/air-technology.

[9] Vasanawala SS, Stormont R, Lindsay S, Grafendorfer T, Cheng JY, Pauly JM, et al. Development and Clinical Implementation of Very Light Weight and Highly Flexible AIR Technology Arrays. *The ISMRM 25th Annual Meeting & Exhibition.* Honolulu, HI, USA; 2017.

[10] Stickle Y-J, Follante C, Giancola M, Anderson D, Robb F, Taracila V, et al. A Novel Ultra-Flexible High-Resolution AIR (Adaptive imaging

receive) 64-Channel Bilateral Phased Array for 3T Brachial Plexus MRI. *ISMRM & SMRT Virtual Conference.* 2020.

[11] Saniour I, Robb F, Taracila V, Vincent J, Voss HH, Kaplitt MG, et al. Acoustically transparent and low-profile head coil for high precision magnetic resonance guided focused ultrasound at 3 T. *International Society for Magnetic Resonance in Medicine.* 2021.

[12] Saniour I, Fraser R, Taracila V, Voss HH, Kaplitt MG, Chazen JL, et al. Attenuation of the dark band artifact in MR-guided focused ultrasound using an ultra-flexible high-sensitivity head coil. *International Society for Magnetic Resonance in Medinice.* 2021.

[13] Hardy CJ, Giaquinto RO, Piel JE, Rohling AAS KW, Marinelli L, Blezek DJ, et al. 128-channel body MRI with a flexible high-density receiver-coil array. *Journal of Magnetic Resonance Imaging: An Official Journal of the International Society for Magnetic Resonance in Medicine* 2008;28(5):1219-25.

[14] Mager D, Peter A, Del Tin L, Fischer E, Smith PJ, Hennig J, et al. An MRI receiver coil produced by inkjet printing directly on to a flexible substrate. IEEE transactions on medical imaging 2010;29(2):482-7.

[15] Jia F, Yuan H, Zhou D, Zhang J, Wang X, Fang J. Knee MRI under varying flexion angles utilizing a flexible flat cable antenna. NMR in Biomedicine 2015;28(4):460-7.

[16] Corea JR, Flynn AM, Lechêne B, Scott G, Reed GD, Shin PJ, et al. Screen-printed flexible MRI receive coils. Nature communications 2016;7(1):1-7.

[17] Frass-Kriegl R, Navarro de Lara LI, Pichler M, Sieg J, Moser E, Windischberger C, et al. Flexible 23-channel coil array for high-resolution magnetic resonance imaging at 3 Tesla. PloS one 2018;13(11):e0206963.

[18] Hosseinnezhadian S, Frass-Kriegl R, Goluch-Roat S, Pichler M, Sieg J, Vít M, et al. A flexible 12-channel transceiver array of transmission line resonators for 7 T MRI. Journal of Magnetic Resonance 2018;296:47-59.

[19] Zhang B, Sodickson DK, Cloos MA. A high-impedance detector-array glove for magnetic resonance imaging of the hand. Nature biomedical engineering 2018;2(8):570-7.

[20] Nohava L, Czerny R, Obermann M, Pichler M, Frass-Kriegl R, Felblinger J, et al. Flexible multi-turn multi-gap coaxial RF coils (MTMG-CCs): design concept and bench validation. *Proceedings of the 27th Annual Meeting of ISMRM, Montreal, Canada.* 2019:0565.

[21] Ruytenberg T, Webb A, Zivkovic I. Shielded-coaxial-cable coils as receive and transceive array elements for 7T human MRI. Magnetic resonance in medicine 2020;83(3):1135-46.

[22] Adriany G, Van De Moortele P-F, Ritter J, Moeller S, Auerbach EJ, Akgün C, et al. A geometrically adjustable 16-channel transmit/receive transmission line array for improved RF efficiency and parallel imaging performance at 7 Tesla. Magnetic Resonance in Medicine 2008;59(3):590-7.

[23] Nordmeyer-Massner JA, De Zanche N, Pruessmann KP. Mechanically adjustable coil array for wrist MRI. Magnetic Resonance in Medicine: An Official Journal of the International Society for Magnetic Resonance in Medicine 2009;61(2):429-38.

[24] Lopez Rios N, Foias A, Lodygensky G, Dehaes M, Cohen-Adad J. Size-adaptable 13-channel receive array for brain MRI in human neonates at 3 T. NMR in Biomedicine 2018;31(8):e3944.

[25] Zhang B, Brown R, Cloos M, Lattanzi R, Sodickson D, Wiggins G. Size-adaptable "Trellis" structure for tailored MRI coil arrays. Magnetic resonance in medicine 2019;81(5):3406-15.

[26] Czerny R, Nohava L, Frass-Kriegl R, Felblinger J, Ginefri J, Laistler E. Flexible multi-turn multi-gap coaxial RF coils: enabling a large range of coil

sizes. *Proceedings of the 27th Annual Meeting of ISMRM, Montreal, Canada.* 2019:1550.

[27] Malko JA, McClees EC, Braun IF, Davis PC, Hoffman J. A flexible mercury-filled surface coil for MR imaging. American journal of neuroradiology 1986;7(2):246-7.

[28] Chitambar CR. Medical applications and toxicities of gallium compounds. International journal of environmental research and public health 2010;7(5):2337-61.

[29] Rumble J, Lide D, Bruno T. CRC handbook of Chemistry and Physics. 2017.

[30] Barta R, Volotovskyy V, Wachowicz K, Fallone BG, De Zanche N. How high can you go? Performance of thin copper and aluminum RF coil conductors. Magnetic Resonance in Medicine 2021;85(4):2327-33.

[31] Edelstein W, Glover G, Hardy C, Redington R. The intrinsic signal-to-noise ratio in NMR imaging. Magnetic resonance in medicine 1986;3(4):604-18.

[32] Zhu S, So J-H, Mays R, Desai S, Barnes WR, Pourdeyhimi B, et al. Ultrastretchable Fibers with Metallic Conductivity Using a Liquid Metal Alloy Core. Advanced Functional Materials 2013;23(18):2308-14.

[33] Mehmann A, Varga M, Vogt C, Port A, Reber J, Marjanovic J, et al. On the bending and stretching of liquid metal receive coils for magnetic resonance imaging. IEEE Transactions on Biomedical Engineering 2018;66(6):1542-8.

[34] Varga M, Mehmann A, Marjanovic J, Reber J, Vogt C, Pruessmann KP, et al. Adsorbed eutectic GaIn structures on a neoprene foam for stretchable MRI coils. Advanced Materials 2017;29(44):1703744.

[35] Mehmann A, Vogt C, Varga M, Port A, Reber J, Marjanovic J, et al. Automatic resonance frequency retuning of stretchable liquid metal receive coil for magnetic resonance imaging. IEEE

transactions on medical imaging 2018;38(6):1420-6.

[36] Port A, Albisetti L, Varga A, Marjanovic J, Reber J, Brunner D, et al. Liquid metal in stretchable tubes: a wearable 4-channel knee array. *Proceedings of the 27th Annual Meeting of ISMRM, Montreal, Canada.* 2019:1114.

[37] Port A, Luechinger R, Albisetti L, Varga M, Marjanovic J, Reber J, et al. Detector clothes for MRi: A wearable array receiver based on liquid metal in elastic tubes. Scientific Reports 2020;10(1):1-10.

[38] Byron K, Winkler SA, Robb F, Vasanawala S, Pauly J, Scott G. An MRI compatible RF MEMs controlled wireless power transfer system. IEEE transactions on microwave theory and techniques 2019;67(5):1717-26.

[39] Port A, Luechinger R, Brunner DO, Pruessmann KP. Conductive Elastomer for Wearable RF Coils. *ISMRM 2020.* 1137. 2020.

[40] Vincent J, Rispoli JV. Conductive thread-based stretchable and flexible radiofrequency coils for magnetic resonance imaging. IEEE Transactions on Biomedical Engineering 2019.

[41] Hayes CE, Axel L. Noise performance of surface coils for magnetic resonance imaging at 1.5 T. Medical physics 1985;12(5):604-7.

[42] Xie J, You X, Huang Y, Ni Z, Wang X, Li X, et al. 3D-printed integrative probeheads for magnetic resonance. Nature communications 2020;11(1):1-11.

[43] Cholleti ER, Stringer J, Assadian M, Battmann V, Bowen C, Aw K. Highly stretchable capacitive sensor with printed carbon black electrodes on barium titanate elastomer composite. Sensors 2019;19(1):42.

[44] Paul CR. Inductance: loop and partial. John Wiley & Sons; 2011.

[45] Fassler A, Majidi C. Soft-matter capacitors and inductors for hyperelastic strain sensing and stretchable

electronics. Smart Mater Struct 2013;22(5).

[46] Igreja R, Dias C. Analytical evaluation of the interdigital electrodes capacitance for a multi-layered structure. Sensors and Actuators A: Physical 2004;112(2-3):291-301.

[47] COMSOL Multiphysics; Available from: https://www.comsol.com.

[48] Ecoflex by Smooth-On; Available from: https://www.smooth-on.com/product-line/ecoflex/.

[49] Sigma-Aldrich, Gallium-Indium eutectic; Available from: https://www.sigmaaldrich.com/US/en/product/aldrich/495425.

[50] Duan Q, Duyn JH, Gudino N, De Zwart JA, Van Gelderen P, Sodickson DK, et al. Characterization of a dielectric phantom for high-field magnetic resonance imaging applications. Medical physics 2014;41(10):102303.

[51] Dielectric phantom recipe generator; Available from: https://amri.ninds.nih.gov/cgi-bin/phantomrecipe. [Accessed 03.03.2021 2021].

# Acknowledgements

This work was supported by the National Institutes of Health under NIH R00EB024341, and GE Healthcare. E.M. thanks Terry Ching of Singapore University of Technology and Design for helpful discussions on material selection and preparation. E.M. and S.A.W. would like to acknowledge Muc Chu, Jojo Borja, and Jonathan Dyke of the Citigroup Biomedical Imaging Center for the helpful technical discussions. Authors thank Cynthia Fox for her valuable assistance in the proofreading of the manuscript.

# Authors contributions

E.M. conceptualized work, performed numerical simulations, built coils, designed experiments, collected data, analyzed data, prepared figures, and wrote the manuscript. E.T.T. conceptualized work, performed imaging experiments and supported data analysis, V.T. conceptualized work and supported data analysis, J.M.V. built coils, T.G. supported data collection, J.S. built coils, H.G.P. conceptualized work, F.J.L.R. conceptualized work and supported data analysis, D.S. conceptualized work, supported data analysis, and provided supervision, S.A.W. conceptualized work, designed experiments, analyzed data, provided supervision and wrote the manuscript. All authors revised the manuscript.